\documentclass{article}
\usepackage{graphicx} 
\usepackage{amsmath, amssymb, amsfonts}
\usepackage{amsthm}%
\usepackage{algorithm}%
\usepackage{algorithmicx}%
\usepackage{algpseudocode}
\usepackage{mathtools}
\usepackage{epstopdf}
\usepackage{textcomp}
\usepackage{multirow}%
\usepackage{xcolor}
\usepackage[mathscr]{euscript}
\DeclareMathOperator*{\argmin}{arg\,min}

\begin{document}
\title{Minimum Attention Control (MAC) in a Receding Horizon Framework with Applications }
\author{Ganesh Teja Theertham,\thanks{Indian Institute of Technology - Hyderabad (IIT-H),Department of Chemical Engineering, IIT-Hyderabad, Kandi, Sangareddy, Telangana - 502285, India, email:ch18m20p000002@iith.ac.in} \,Santhosh Kumar Varanasi,\thanks{Indian Institute of Technology - Jodhpur (IIT-J),Department of Chemical Engineering, IIT-Jodhpur,\\ Nagaur Road, Karwar, Jodhpur, Rajasthan - 342037, India, email:skvaranasi@iitj.ac.in}\,Phanindra Jampana,\thanks{Indian Institute of Technology - Hyderabad (IIT-H),Department of Chemical Engineering, IIT-Hyderabad, Kandi,Sangareddy, Telangana - 502285, India, email:pjampana@che.iith.ac.in}}
\date{\today}

\maketitle
\begin{abstract}
Minimum Attention Control (MAC) is a control technique that provides minimal input changes to meet the control objective. 
Mathematically, the zero norm of the input changes is used as a constraint for the given control objective and minimized 
with respect to the process dynamics. In this paper, along with the zero norm constraint,  
stage costs are also considered for reference tracking in a receding horizon framework. For this purpose, the 
optimal inputs of the previous 
horizons are also considered in the optimization problem of the current horizon. An alternating minimization 
algorithm is 
applied to solve the
optimization problem (Minimum Attention Model Predictive Control (MAMPC)). The outer step of the optimization is 
a quadratic program, while the inner step, which solves for sparsity, has an analytical solution. The proposed algorithm is 
implemented on two case studies: a four-tank system with slow dynamics and a fuel cell stack with fast dynamics. A 
detailed comparative study of the proposed algorithm with standard MPC indicates sparse control actions with a tradeoff 
in the tracking error. 
\end{abstract}

\section{Introduction}
\label{sec:introduction}
Sparsity in control actions is desirable to reduce the wear and tear of the actuating elements. Techniques such as Minimum Attention Control (MAC) 
~\cite{657776, nagahara2020approach} and Maximum Hands Off control 
(MHC)~\cite{nagahara2016discrete,nagahara2015maximum} 
enforce sparsity in the control actions. The practical benefits of sparsity in control are significant in 
real-world scenarios. MHC is used in a wide range of applications such as automobiles, railway vehicles, and embedded 
systems~\cite{nagahara2020sparsity}. For example, in automobiles, MHC shuts off the engine during the idle period, 
reducing greenhouse gas emissions. Hence, MHC is also popularly called green control\cite{nagahara2016discrete}. 

In MAC, the changes in the control actions are minimized, whereas in the MHC, the control action itself is made negligible 
for long periods. In other words, MHC is the control that possesses the smallest support over all the feasible controls 
that drive the state to zero at a fixed final time. In MAC, the input change rate is considered in the objective function, and the 
optimization is carried over all inputs, which takes the state to zero at a fixed final time. This results in a control 
signal with minimal input changes. The explicit usage of sparse constraint either on the input or input change is not considered in self-triggered 
and event-triggered control~\cite{heemels2012introduction}, since the purpose of aperiodic 
control is to find the control execution time steps using the triggering conditions obtained from the performance of the 
closed loop. Donkers et.al~\cite{donkers2011minimum,donkers2014minimum} consider MAC 
for linear systems where the discrete-time control inputs $u_k,\; k \in \mathrm{N}$ and 
the sequence of execution instants $\{t_k\}_{k\in \mathrm{N}}$ are given by control problem formulated using a linear 
program. For non-linear systems, the existence of MAC solutions is not always 
assured. In Pilhwa Lee et al.\cite{lee2021existence} for a class of nonlinear systems where the input $u(t)$ 
contains a feedback gain and a feedforward term, an optimal solution to the MAC was 
shown to exist.

For MAC, Nagahara et al. ~\cite{nagahara2020approach} consider continuous time linear dynamical systems and show that the 
optimal control obtained using $L_1$-relaxation is also optimal for $L_0$-norm provided certain conditions are met. 
The 
objective function considered in their paper is $\|\frac{du}{dt}\|_0$, where the norm is defined as the Lebesgue 
measure of the support of the derivative of $u$. Similarly, Nagahara et al. ~\cite{nagahara2015maximum} provide guarantees for MHC that 
the $L_1$-relaxation also provides solutions to the $L_0$ problem, which consists of minimizing $\|u\|_0$, provided
certain conditions are satisfied. In both these papers, the objective is to bring the state of the system to zero, i.e. 
the regulation problem.
During control design, important factors, such as constraints related to the quality of the 
product and safety limitations from the system, need to be accounted for. Therefore, in this paper, we consider more 
practical objective functions which 
also involve reference tracking. 

In the discrete case too~\cite{nagahara2016discrete}, $l_0$-norm provides significant computational challenges owing to 
its non-convex and combinatorial nature. Therefore, a popular and widely used technique is replacing the $l_0$-norm with 
the relaxed convex version, i.e., the $l_1$ norm. Convex optimization techniques can then be used to solve the resulting 
optimization problem efficiently. Apart from $l_1$ relaxations, the sparse constraint can also be enforced using non-convex optimization approaches~\cite{ikeda2024non}, in particular, non-convex penalty functions such as smoothly clipped absolute deviation (SCAD), minimax concave penalty (MCP) and log sum penalty (LCP) are considered to be surrogates for the $l_0$ constraint. In classical control strategies such as Linear 
Quadratic Regulator (LQR), the standard quadratic cost, can also be augmented with a $l_1$ norm or the total variation, 
resulting in control signals that are either sparse or infrequently change when we march forward in time. The resulting 
control design is named Sparse Quadratic 
Regulator~\cite{6669833}.

A combined $L^1$-$L^2$ norm (CLOT-norm) is used to achieve a much sparser control than $L^1/L^2$ 
optimal control, which is 
continuous and unique~\cite{nagahara2020clot}. The theoretical results are based on Pontryagin's 
minimum 
principle ~\cite{athans2013optimal}. Motivated by the MAC, a control algorithm for systems described by stochastic 
differential equations is developed in Varanasi et.al~\cite{varanasi2021minimum}. In a simulation 
case study on a 
quadruple tank system, it has been noted that this control obtained has a small number of 
changes~\cite{varanasi2021minimum}. 

A sparsifying control called $lasso$-MPC has been proposed
in~\cite{gallieri2012} which employs the $l_1$-norm on the control action in a receding horizon 
framework. 
An MPC with $l_0$ constraint is considered in~\cite{aguilera2014quadratic} where the zero norm 
of the input is constrained at every time instant in the horizon.
The zero norm constraint is 
enforced by writing 
the sum of the smallest elements of a vector as a minimization problem. The overall optimization is 
performed 
iteratively with inner and outer steps, which are partially performed over non-negative orthants. 

In the current paper, the objective function contains a quadratic penalty for output and input changes 
with a zero norm constraint for the input changes. The current paper also considers alternating optimization with 
inner and outer steps. This alternating
optimization technique has also been applied in~\cite{varanasi2021minimum}, even though the authors do not 
consider the receding horizon approach. The differences between the current paper and the previous papers are 
twofold: 1) 
We consider the zero norm on the input change instead of the input as considered in ~\cite{gallieri2012} and 
~\cite{aguilera2014quadratic} 2) The inner step has 
an analytical solution given by the best sparse approximation in contrast 
to~\cite{aguilera2014quadratic}. Therefore, the resulting optimization will be faster. Further, considering only the zero 
norm on the input does not require knowledge of the previous optimal input in the current 
horizon. However, in the present paper, since the zero norm on the input changes is considered, this knowledge is also 
accounted for in the optimization.

The paper is organized as follows. In Sec.\ref{sec:formulation}, the mathematical formulation of the Minimum Attention 
Model Predictive Control (MAMPC) problem is provided. The alternating algorithm to obtain an approximate solution 
to the MAMPC problem is given in Sec.~\ref{sec:algorithm}. The performance of the 
alternating algorithm is demonstrated with two simulation 
case studies: Quadruple Tank System, which has slow dynamics, and Solid Oxide Fuel Cell (SOFC) stack, which has fast 
dynamics 
and results are discussed in Sec.~\ref{sec:numericalsim}. Finally, concluding remarks are given in 
Sec.~\ref{sec:conclusion}.

\section{Formulation of the MAMPC problem}
\label{sec:formulation}
Consider a discrete time state space model of the form
\begin{equation*}
\begin{split}
    x_{k+1} &= Ax_k +Bu_k\\
    y_k &= Cx_k + Du_k
\end{split}
\label{eq:sys_dyn}
\tag{1}
\end{equation*}
 where $u_k \in \mathrm{R}^m, y_k \in \mathrm{R}^l$ and $x_k \in \mathrm{R}^{n}$ are input, output and state vectors 
 respectively. System matrices $A,B,C,D$, which govern the state and observation dynamics, have appropriate sizes 
 according to the dimensions of $u_k,y_k$, and $x_k$, respectively. 

Let $n_p$ and $n_c$ be the prediction and the control horizons, respectively. In MPC, using the model, the future states 
are predicted until the prediction horizon $(n_p)$ and an optimization problem is solved to find the optimal input 
$\mathbf{u}$ which minimizes the $l_2$-norm of the error subject to system dynamics and constraints on $\mathbf{x}$ and 
inputs $\mathbf{u}$ over the input horizon $(n_c)$.

Let $\mathbf{x_k} \coloneq \{x_{k+1},\hdots \hdots x_{k+n_p}\}$ be the collection of all the states from $k+1$ to 
$k+n_p$. In a similar manner, let $\mathbf{u_k} \coloneq \{u_k,u_{k+1},\hdots \hdots u_{k+n_c-1}\}$ be the collection of 
all the inputs from $k$ to $k+n_c-1$. The reference signal for the control objective and the successive 
difference or change in the input vector are denoted by $r_k$ and $\Delta u_{k+j} \coloneq u_{k+j} - u_{k+j-1}$ respectively.
The
objective or performance measure in MPC is given by
 \begin{equation*}
     J_k(\mathbf{u_k},\mathbf{y}_k) \coloneq  \sum_{i=0}^{n_p} c_1(r_{k+i},y_{k+i},u_{k+i})+ c_2(r_{k+n_p},y_{k+n_p})
     \label{eq:mpc_obj}
     \tag{2}
 \end{equation*}
 Where $c_1,c_2$ denotes the stage and terminal costs. In this paper, we consider the input changes in the 
 stage cost, i.e., the objective function is considered to be
 \begin{equation*}
     J_k^{mpc}(\mathbf{u_k},\mathbf{y}_k) = \sum_{i=0}^{n_p} \|r_{k+i} -y_{k+i}\|_2^2 + \lambda \sum_{j=0}^{n_c-1} \Delta u_{k+j}^2\\
     \tag{3}
 \end{equation*}
In MPC, the above objective is minimized with respect to the system dynamics.
It can be noted that 
the input change penalty term is introduced to restrict large changes in the input. However, 
this does not avoid small and frequent changes in the input, which can result in wear and tear
of the control valves. To reduce frequent changes in the input, a sparse constraint, i.e., a 
constraint on the $l_0$-norm of the input changes, is enforced in the receding horizon 
framework. The formulation for this problem called the Minimum Attention Model Predictive 
Control (MAMPC) is discussed in the following.

Let $n_s \geq 0$ be the 
sparsity horizon, $\tilde{x}$ be the initial state given at the timestep $k$ and $u_k^{n_s}$ be input vector when the past inputs are included, i.e.,
$$\mathbf{u}_k^{n_s} \coloneqq [ u_{k-n_s} u_{k-n_s+1} u_{k-n_s+2} \hdots u_{k},u_{k+1} \hdots u_{k+n_c-1}]$$

Then, the MAMPC problem is defined as follows.
 \begin{equation*}
 \begin{split}
          \argmin_{u_{k-n_s : k+n_c-1}} \quad &\sum_{i=0}^{n_p} \|r_{k+i} -y_{k+i}\|_2^2 + \lambda \sum_{j=1}^{n_c-1} \|\Delta u_{k+j}\|^2_2 \\
      \text{s.t } \quad x_k &= \Tilde{x}\\
    x_{k+j} &= Ax_{k+j-1} + Bu_{k+j-1},\,j = 1,2,\hdots,n_c\\
    y_{k+j-1} &= Cx_{k+j-1} + Du_{k+j-1},\, j = 1,2,\hdots,n_c\\
    x_{k+i} &= Ax_{k+i-1} + Bu_{k+n_c-1},\,i = n_c+1,\hdots,n_p\\
    y_{k+i} &= Cx_{k+i} + Du_{k+n_c-1},\, i = n_c,\hdots,n_p\\
    u_{min} &\leq u_{k+j} \leq u_{max},\,j = 0,1,2,\hdots,n_c-1\\
    y_{min} &\leq y_{k+j} \leq y_{max},\,j = 0,1,2,\hdots,n_p \\
    u_{k-l} &= u_{k-l}^*,\, l = 1,2,\cdots,n_s\\
   &\sum_{i=-n_s+1}^{n_c-1} \|u_{k+i} - u_{k+i-1}\|_{0} \leq s  \\
\end{split}
\label{eq:mampcocp}
\tag{$P_0$}
 \end{equation*}
 where $s$ is the predefined level of sparsity by which the number of input changes can be 
 bounded in the horizon and $u_{k-l}^*, l = 1,2,\cdots,n_s$ are the optimum inputs from the previous 
 horizon.
 The zero norm constraint can be written appropriately  by defining the matrix 
 $\boldsymbol{\Psi}$ as follows.
 $$\|\boldsymbol{\Psi} \boldsymbol{\upsilon}\|_0 \leq s$$
where $\boldsymbol{\Psi}$ is a block 
triangular matrix which converts the input 
vector to successive difference of the 
inputs and $\boldsymbol{\upsilon} = 
\text{vec}((\mathbf{u}_k^{n_s})^T)$, a 
$(n_c+n_s)m \times 1$ vector. 
$\boldsymbol{\Psi}$ is given by
\begin{equation*}
\boldsymbol{\Psi} =
\begin{bmatrix}
     \bar{\aleph} & \bar{0}  & \bar{0} 
     &\cdots  & \bar{0} \\
     \bar{0} & \bar{\aleph} & \bar{0} 
     &\cdots  & \bar{0} \\
     \vdots &  \vdots & \ddots & \ddots & 
     \vdots\\
     \bar{0} &\bar{0} & \cdots & \cdots & 
     \bar{\aleph}\\
\end{bmatrix}
\end{equation*} where $\bar{\aleph}$ is a upper bi-diagonal matrix which contains $-1$ on the main diagonal and $1$ 
on the super diagonal as shown below. The size of the $\bar{\aleph}$ and $\bar{0}$ is $(n_c+n_s-1) \times 
(n_c+n_s)$. Therefore, $\boldsymbol{\Psi}$ is a matrix of size $m(n_c+n_s-1) \times m(n_c+n_s)$ and 
\begin{equation*}
    \bar{\aleph} = 
    \begin{bmatrix}
        -1 & 1 & 0 & \hdots & 0 & 0\\
        0 & -1 & 1 & \hdots & 0 & 0 \\
        \vdots & \vdots & \ddots & \ddots  & \vdots & \vdots\\
        0 & 0 & 0 & \hdots  & -1 & 1\\
    \end{bmatrix}
\end{equation*}
 
 The above ($P_0$) problem involves quadratic optimization with respect to the $l_0$ norm constraint. Since the 
 zero-norm constraint is non-convex, the optimization problem cannot be solved efficiently. Therefore, the above 
 problem is
 reformulated as given below. Denoting $\boldsymbol{\Psi} \boldsymbol{\upsilon} = \boldsymbol{\hat{\upsilon}}$ and 
 by 
 introducing Lagrange multiplier $\mu$, the following problem is posed. 
 \begin{equation*}
     \begin{split}
         \argmin_{\boldsymbol{\upsilon},\boldsymbol{\hat{\upsilon}},\|\boldsymbol{\hat{\upsilon}}\|_0 \leq s} \quad &J_k^{mampc}(\mathbf{y}_k,\boldsymbol{\upsilon},\boldsymbol{\hat{\upsilon}}) := \sum_{i=0}^{n_p} \|r_{k+i} -y_{k+i}\|_2^2 + \lambda \sum_{j=1}^{n_c-1} \|\Delta u_{k+j}\|^2_2 + \mu \| \boldsymbol{\hat{\upsilon}}-\boldsymbol{\Psi} \boldsymbol{\upsilon} \|_2^2 \\
      \text{s.t } \quad x_k &= \Tilde{x} \\
    x_{k+j} &= Ax_{k+j-1} + Bu_{k+j-1},\,j = 1,2,\hdots,n_c\\
    y_{k+j-1} &= Cx_{k+j-1} + Du_{k+j-1},\, j = 1,2,\hdots,n_c\\
    x_{k+i} &= Ax_{k+i-1} + Bu_{k+n_c-1},\,i = n_c+1,\hdots,n_p\\
    y_{k+i} &= Cx_{k+i} + Du_{k+n_c-1},\, i = n_c,\hdots,n_p\\
     u_{min} &\leq u_{k+h} \leq u_{max},\,h = 0,1,2,\hdots,n_p-1\\
    y_{min} &\leq y_{k+h} \leq y_{max},\,h = 0,1,2,\hdots,n_p\\
    u_{k+l} &= u_{k+l}^*,\, l = -n_s,\hdots,-1\\
     \end{split}
     \label{eq:mampcform}
     \tag{$P_1$}
 \end{equation*}
    
 In the reformulation, a new variable $\boldsymbol{\hat{\upsilon}}$ is introduced, which serves as 
 an optimization variable 
 for the zero norm constraint. It can be noted that any feasible input to $(P_0)$ is also 
 a feasible input to $(P_1)$. 
 However, the reverse may not be true. Further, the optimal value of $(P_1)$ is less or equal to 
 the
 optimal value of $(P_0)$. In order to solve the above problem, instead of joint minimization of 
 $\mathbf{u}_k^{n_s},\boldsymbol{\hat{\upsilon}}$ an 
 alternating minimization technique is followed. 

 \section{Alternating Minimization Algorithm for MAMPC}
 \label{sec:algorithm}
 This section provides an alternating minimization algorithm to solve the optimization problem \ref{eq:mampcform}, which involves two steps. 
 In the first step, the following optimization problem is solved. An initial guess of 
 $\boldsymbol{\hat{\upsilon}}$ is used to obtain optimal $\boldsymbol{\upsilon}^*$.

\begin{align*}
    \argmin_{\boldsymbol{\upsilon}} \quad & J_k^{mampc}(\mathbf{y}_k,\boldsymbol{\upsilon},\boldsymbol{\hat{\upsilon}^*}) \label{eq:6} \tag{6} \\
      \text{s.t } \quad x_k &= \Tilde{x}\label{eq:6a} \tag{6a}\\
    x_{k+j} &= Ax_{k+j-1} + Bu_{k+j-1},\,j = 1,2,\hdots,n_c \label{eq:6b} \tag{6b}\\
    y_{k+j-1} &= Cx_{k+j-1} + Du_{k+j-1},\, j = 1,2,\hdots,n_c \label{eq:6c}\tag{6c}\\
    x_{k+i} &= Ax_{k+i-1} + Bu_{k+n_c-1},\,i = n_c+1,\hdots,n_p \label{eq:6d}\tag{6d}\\
    y_{k+i-1} &= Cx_{k+i-1} + Du_{k+n_c-1},\, i = n_c+1,\hdots,n_p+1 \label{eq:6e}\tag{6e}\\
    u_{min} &\leq u_{k+j} \leq u_{max},\,h = 1,2,\hdots,n_p-1 \label{eq:6f}\tag{6f}\\
    y_{min} &\leq y_{k+j} \leq y_{max},\,h = 1,2,\hdots,n_p \label{eq:6g}\tag{6g}\\
    u_{k+l} &= u_{k+l}^*,\, l = -n_s,\hdots,-1\label{eq:6h}\tag{6h}\\
\end{align*}

In the second step, the optimal solution $\boldsymbol{\upsilon}^*$ is used to obtain the optimal  
$\boldsymbol{\hat{\upsilon}}^*$ and 
the process is repeated in an alternating manner.

The optimization problem in the second step has an explicit solution. In the objective function, only the third term is 
involved, the other two terms can be dropped since the optimization is with respect to $\boldsymbol{\hat{\upsilon}}$. 
The optimization in the second step can be simplified as follows.

\begin{align*}
    \argmin_{\boldsymbol{\hat{\upsilon}}}& \quad \| \boldsymbol{\hat{\upsilon}}-\boldsymbol{\Psi} \boldsymbol{\upsilon}^*\|_2^2 \label{eq:7} \tag{7}\\
    \text{s.t } & \quad\|\boldsymbol{\hat{\upsilon}}\|_0 \leq s\label{eq:7a} \tag{7a}\\
\end{align*}
The optimal solution (also known as the best $r$-sparse solution) can be obtained by taking the $s$ maximum components of the absolute values of the 
vector $\boldsymbol{\Psi} \boldsymbol{\upsilon}^*$. The optimization problem $(P_1)$ is solved in an alternating approach to obtain 
$\boldsymbol{\upsilon}^*, \boldsymbol{\hat{\upsilon}}^* $ with a pre-specified error tolerance.
The alternating minimization may not solve $(P_1)$. Further, as 
noted before, $(P_1)$ may also result in a solution that is different from $(P_0)$.
However, we 
show with the help of numerical simulations in the next section that the resulting control 
reduces the changes in the inputs without significantly sacrificing the tracking accuracy. The overall 
optimization method is described in Algorithm~\ref{alg:alternating_min}.

 \begin{algorithm}
   \caption{Alternating Minimization algorithm for  Minimum Attention Model Predictive Control (MAMPC)}
   \label{alg:alternating_min}
  \begin{algorithmic}[]
     \Require System : Reference signal $(y_{ref})$, Initial state $(x_0)$, Final timestep $(N)$, Pre-specified error tolerance $(\epsilon_1)$
     \Ensure Prediction horizon $(n_p)$, Control horizon $(n_c)$, Sparsity horizon $(n_s)$, Weight Matrices $\{Q,R\}$, Sparsity level $(r)$
     \State Initialization with a guess value of $\boldsymbol{\hat{\upsilon}}$
     \While{$k \leq N$}
     \State iteration $i = 0$
     \While{$\|\boldsymbol{\upsilon}^{*}_{i+1}-\boldsymbol{\upsilon}^{*}_{i}\|_1 \leq \epsilon_1$}
     \State \textbf{First Step}
     \State Minimize the objective $J_k^{mampc}(\mathbf{y}_k,\boldsymbol{\upsilon},\boldsymbol{\hat{\upsilon}^*})$ w.r.t $\boldsymbol{\upsilon}$
     \State \textbf{s.t}
     \State \quad Eq~\eqref{eq:6a}-\eqref{eq:6e} System constraints (State and Measurement Dynamics) 
     \State \quad Eq~\eqref{eq:6f}-\eqref{eq:6g} Inequality constraints (Limitations of inputs and measurements)
     \State \textbf{Second Step}
     \State Using the optimal solution $\boldsymbol{\upsilon}^*_k$ obtained from \textbf{First step}, Minimize $J_k^{mampc}(\mathbf{y}_k,\boldsymbol{\upsilon}^*_k,\boldsymbol{\hat{\upsilon}})$ w.r.t $\boldsymbol{\hat{\upsilon}}$
     \State \textbf{s.t}
     \State \quad Eq~\eqref{eq:7a} Sparse constraint on successive difference of input vector
     \State Explicit solution can be given for the above optimization problem.
     \State Using solution  $\boldsymbol{\hat{\upsilon}^*}$,\quad $i=i+1$ go to \textbf{First Step}
     \EndWhile 
     \EndWhile
  \end{algorithmic}
  \end{algorithm}
\section{Numerical Examples}
\label{sec:numericalsim}
This section presents numerical simulations of the alternating algorithm on a Quadruple Tank System and a Fuel Cell System. A comparison with MPC is also provided.

\subsection{Case study $1$ : Quadruple Tank System}
The control inputs obtained from the alternating algorithm are implemented on the quadruple tank setup, and the results are compared with those of MPC.
\begin{figure}[h!]
    \centering \includegraphics{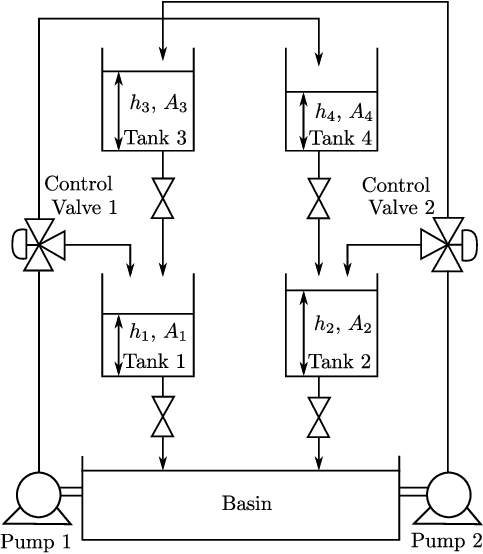}
    \caption{Quadruple tank system where Tanks $1,3$ and Tanks $2,4$ are configured in non-interacting manner.}
    \label{fig:scfour}
\end{figure}
The governing equations for the quadruple tank system are considered from literature~\cite{johansson2000quadruple},
\begin{equation*}
\begin{split}
    \frac{dh_1}{dt} &= -\frac{a_1}{A_1}\sqrt{2gh_1} + \frac{a_3}{A_1}\sqrt{2gh_3} + \frac{\gamma_1}{A_1}f_1\\ 
    \frac{dh_2}{dt} &= -\frac{a_2}{A_2}\sqrt{2gh_2} + \frac{a_4}{A_2}\sqrt{2gh_4} + \frac{\gamma_2}{A_2}f_2\\
    \frac{dh_3}{dt} &= -\frac{a_3}{A_3}\sqrt{2gh_3} + \frac{1-\gamma_2}{A_3}f_2\\
     \frac{dh_4}{dt} &= -\frac{a_4}{A_4}\sqrt{2gh_4} + \frac{1-\gamma_1}{A_4}f_1
\end{split}
\label{eq:fourtankeqs}
\tag{8}
\end{equation*}
Where $A_i, a_i, h_i$ corresponds to cross-sectional area, outlet cross-sectional area, and level of the tank, respectively. $f_i = k_i\nu_i$ where $k_i$ is the pump proportionality constant and $\nu_i$ is the control valve opening, which is considered as a manipulated variable instructed by the user. The water flow from each of the control valves (CV) splits with a specific ratio, which is known as $\gamma_1,\gamma_2$.
 $\gamma_1$ is the ratio of flow in Tank-$1$ to total flow from the control valve $1$. Similarly, $ \gamma_2$ is the ratio of flow in Tank-$2$ to total flow from the control valve $2$.
 $\gamma_1,\gamma_2$ govern the zero location of the system. If the sum $\sum_{i=1}^2 \gamma_i < 1$, then the plant is under minimum phase dynamics. Otherwise, the dynamics of the plant are a non-minimum phase.
It is assumed that only the first two states are measured, and hence, the observation model is given by
\begin{equation*}
    y_k = k_c\begin{bmatrix}
        1 & 0 & 0 & 0\\
        0 & 1 & 0 & 0
    \end{bmatrix} x_k\\
    \label{eq:fourtankobseq}
    \tag{9}
\end{equation*}
\begin{table}[h!]
    \centering
    \begin{tabular}{|l|c|c|c|}
      \hline
      Parameters   &  Symbol & Unit & Value \\
      \hline
      Cross-sectional area of tank $i$   &  $A_i$ & $cm^2$ & 730\\
      Outlet cross-sectional area of tank $i$ & $a_i$ & $cm^2$ & 2.05,2.26,2.37,2.07\\
      Gravitational constant & $g$ & $cm/s^2$ & 981\\
      Level sensor calibration constant & $k_c$ & & 2\\
      Ratio of flow in Tank-1 to total flow from CV-1 & $\gamma_1$ & & 0.3\\
      Ratio of flow in Tank-2 to total flow from CV-2 & $\gamma_2$ & & 0.3\\
      \hline
    \end{tabular}
    \caption{Physical constants in quadruple tank system~\cite{varanasi2021minimum}}
    \label{tab:exp_values}
\end{table}

\subsubsection{Simulation}
The above continuous time model is simulated to obtain the input-output data until the steady state is achieved. The
input values for this steady state can be referred from the Table~\ref{tab:ss_vals}.
\begin{table}[h!]
    \centering
    \begin{tabular}{|l|c|c|}
        \hline
        Signal & Symbol & Values\\
        \hline
        Input values & CV-1, CV-2 & 50,50 \\
        Output values & $h_1$,$h_2$,$h_3$,$h_4$  & 16.3,13.7,6.0,8.1\\
        \hline
    \end{tabular}
    \caption{Steady state values of the quadruple tank system}
    \label{tab:ss_vals}
\end{table}

A pseudo random binary sequence (PRBS) signal~\cite{ljung1998system} is considered as input. In the simulation, we hold the input value for a specific time, during which the effect of the input can be 
observed. A magnitude of $25$ percent is used as a perturbation signal. Therefore, PRBS signal magnitude switches around 
the steady state value i.e, $50\pm 25\%$. The order of the PRBS signal is considered to be 8. Since the plant consists of 
$2$ inputs, a different initial bit sequence is used to generate a PRBS signal for each of the inputs. The plant is 
simulated in MATLAB-Simulink\textsuperscript{\textregistered}, and the input-output data is used for the subspace identification exercise.
 
\begin{figure}[h!]
    \centering
    \includegraphics[width=0.5\linewidth]{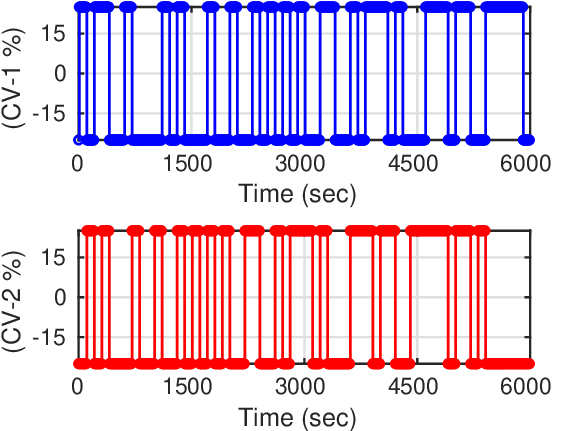}
    \caption{PRBS as input for each of the control valves (CV) in quadruple tank system}
    \label{fig:u_prbs}
\end{figure}

\begin{figure}[h!]
    \centering
    \includegraphics[width=0.5\linewidth]{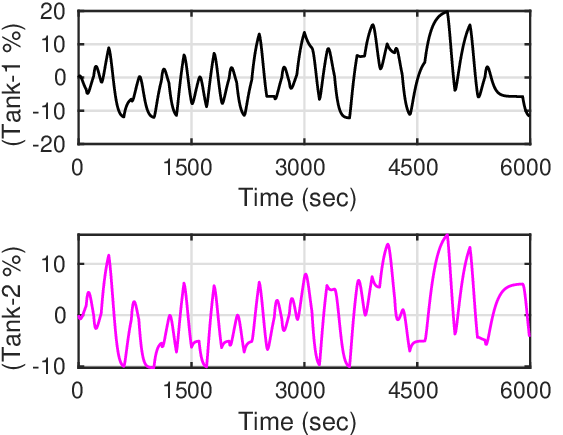}
    \caption{Levels of tanks-1 \& 2 in quadruple tank system for the PRBS inputs in Fig.~\ref{fig:u_prbs}}
    \label{fig:y_tanks}
\end{figure}

For the identification exercise, it is assumed that the level is measured only in tanks-$1,2$. The input-output data 
after the preprocessing step are shown in Fig.~\ref{fig:u_prbs},~\ref{fig:y_tanks}. Only a part of the full-length 
PRBS is shown in the above-mentioned figures for better visualization.

\subsubsection{ Identification}
The SYSID toolbox is used to identify a linear discrete-time state space model. The standard subspace identification 
technique (N4SID) algorithm \cite{tangirala2018principles,van1994n4sid} has been used to obtain state space matrices.

\subsubsection{Comparison of MPC and MAMPC Results}
\begin{figure}[h!]
    \centering
    \includegraphics[width=0.5\linewidth]{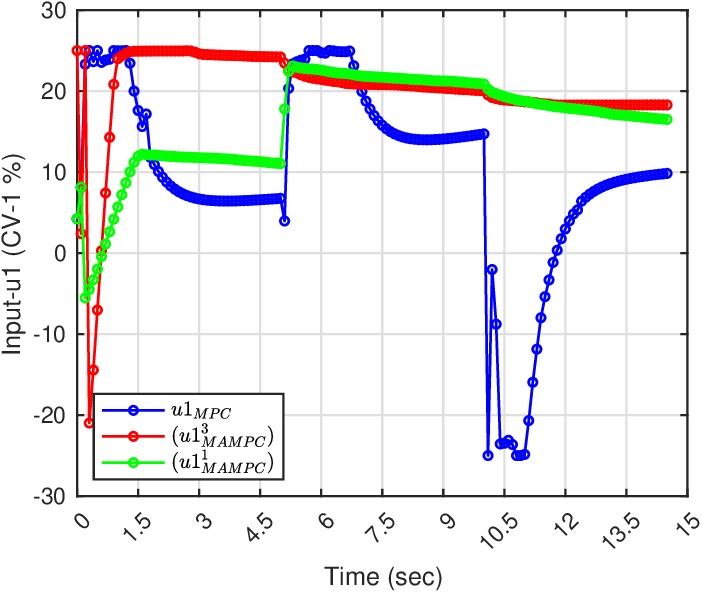}
    \caption{Optimal profiles of manipulated variable control valve-1 (CV1 in $\%$)  for the quadruple tank system given by MPC and MAMPC (two cases are considered with $n_s=1$ and $n_s=3$ for a fixed sparsity level $s = 3$)}
    \label{fig:q_tank_u1_all}
\end{figure}

\begin{figure}[h!]
    \centering
    \includegraphics[width=0.5\linewidth]{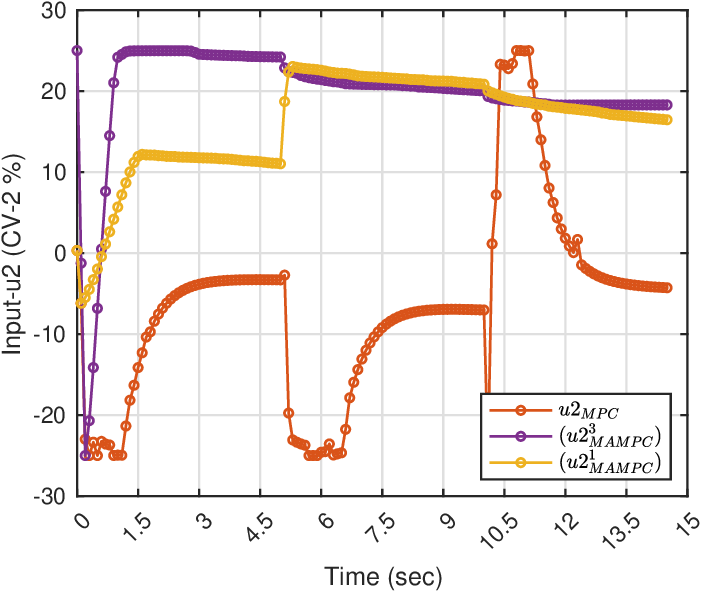}
    \caption{Optimal profiles of manipulated variable control valve-2 (CV2 in $\%$)  for the quadruple tank system given by MPC and MAMPC (two cases are considered with $n_s=1$ and $n_s=3$ as before and $s = 3$)}
    \label{fig:q_tank_u2_all}
\end{figure}

The alternating minimization algorithm~Algorithm~\ref{alg:alternating_min} is implemented to 
solve MAMPC 
problem using the CVX toolbox in MATLAB with prediction horizon $(n_p = 10)$, control horizon 
$(n_c = 5) $ and the sparsity horizon  $(n_s = 1,3)$. The optimal profiles for MPC and MAMPC 
with the design parameters mentioned above for the quadruple tank system are shown in 
Figures~\ref{fig:q_tank_u1_all},\ref{fig:q_tank_u2_all}. 

It is to be noted that the proposed method requires past input data to enforce the sparse constraint in the 
optimization problem. Hence, in the first few steps, MPC is implemented on the quadruple tank system and switched to
the alternating minimization algorithm to solve the MAMPC problem.

It is evident from Figure~\ref{fig:q_tank_u1_all} for the input component$(u1)$, $u1^3_{MAMPC}$ does 
not change frequently compared to $u1_{MPC}$ and $u1^1_{MAMPC}$.  In $u1_{MPC}$, the input fluctuations can be seen specifically when the reference 
signal changes to another value. In the case of $u1^3_{MAMPC}$ and $u1^1_{MAMPC}$, the magnitude of changes
is much smaller.
For input component$(u2)$, $u2^3_{MAMPC}$ has minimal input changes compared to $u1_{MPC}$ and $u1^1_{MAMPC}$ as 
shown in Fig.~\ref{fig:q_tank_u2_all}. In fact, the optimal profiles of $u1$ and $u2$ are very similar in the case of MAMPC.

\begin{figure}[h!]
    \centering
    \includegraphics[width=0.5\linewidth]{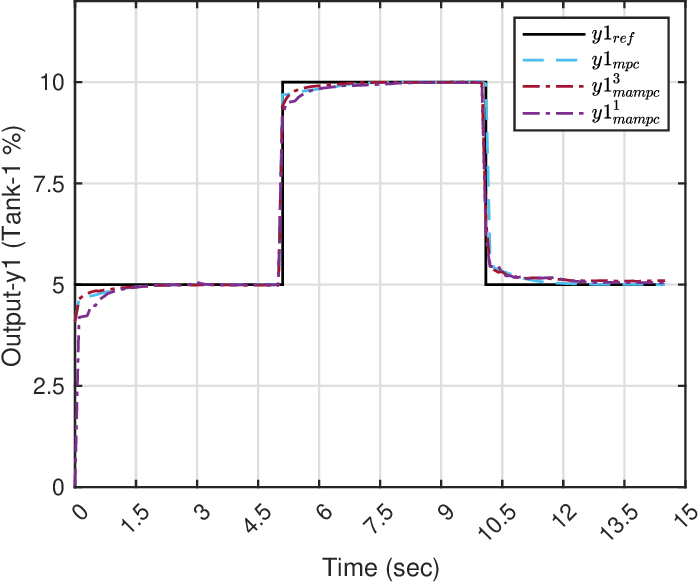}
    \caption{Output level of tank - $1$ for the given reference in quadruple tank system for optimal input profiles given by MPC and MAMPC (two cases are considered with $n_s=1$ and $n_s=3$ as before and $s = 3$)}
    \label{fig:q_tank_y1_all}
\end{figure}

\begin{figure}[h!]
    \centering
    \includegraphics[width=0.5\linewidth]{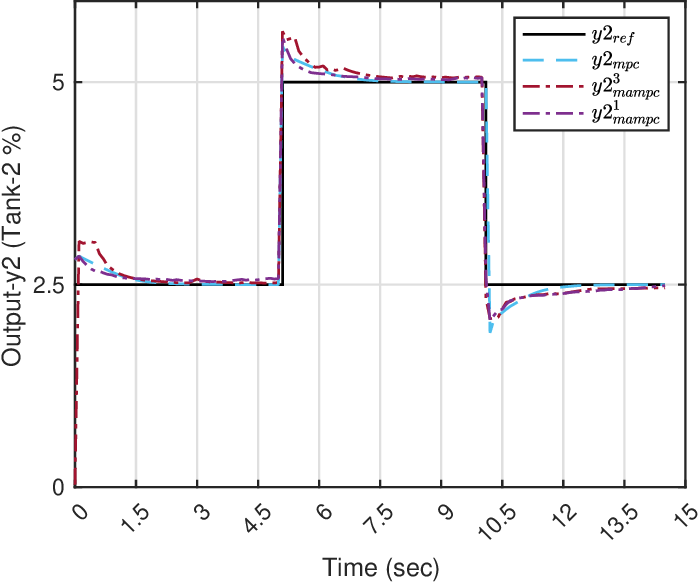}
    \caption{Output level of tank - $2$ for the given reference in quadruple tank system for optimal input profiles given by MPC and MAMPC (two cases are considered with $n_s=1$ and $n_s=3$ as before and $s = 3$)}
    \label{fig:q_tank_y2_all}
\end{figure}

The closed-loop response of the quadruple tank system is shown in Figures~\ref{fig:q_tank_y1_all},~\ref{fig:q_tank_y2_all}. In view of tracking performance, the optimal responses are satisfactory for MPC and MAMPC. The Mean Square Error (MSE) for tracking is given as
\begin{equation*}
    \text{Tracking Error } (\epsilon) = \sum_{i=1}^n \frac{1}{n} \|y_i-y_{{ref}_i}\|_2^2
    \label{eq:trackerror}
    \tag{11}
\end{equation*}
Tracking errors for each of the responses are reported in the Table~\ref{tab:sd_te_q_tank_all}.

In 
literature~\cite{nagahara2020clot} the sparse density defined by $\frac{\|u\|_0}{T}$ is used to measure how often the 
control input is active for a given time horizon. In the current work, to measure how often the input changes, we rely 
on the above-mentioned definition, but calculate sparse density on the successive difference of the input  
defined as 
\begin{equation*}
    \text{Sparse density }(\vartheta) = \frac{\|\Delta u\|_0}{\#\left[0,T\right]}
    \label{eq:sd}
    \tag{10}
\end{equation*}
Where $\#\left[0, T\right]$ is the number of time steps considered in the given interval. This allows us to compare the 
control inputs from the MPC and MAMPC in an effective way.
For the MIMO system considered, we calculate the sparse density for each of the components of input with a threshold on the input change.
The following table compares both the optimal control inputs using sparse density as defined above. It is to be noted
that the changes in the input are not exactly zero, even for the solution obtained from the proposed algorithm and, therefore, in the 
calculation of the sparse density, a threshold is applied to each component of the input difference.

\begin{table}[]
    \centering
    \begin{tabular}{|c|c|c|c|c|}
    \hline
       \multicolumn{1}{|p{2cm}|}{Method} & \multicolumn{1}{|p{2cm}|}{Sparse density $(\vartheta_{u_1})$} & \multicolumn{1}{|p{2cm}|}{Sparse density $(\vartheta_{u_2})$} &  \multicolumn{1}{|p{2cm}|} {Tracking error $(\epsilon)$}\\
        \hline
        MAMPC ($n_s=1$) & 0.241 & 0.241  & 0.048 \\
        MAMPC ($n_s=3$) & 0.193 & 0.193  & 0.257\\
        MPC & 0.579 & 0.613  & 0.217\\
        \hline
    \end{tabular}
    \caption{Quadruple tank system : Comparison of sparse density for a threshold value of $0.1$ and tracking errors (for a fixed value of $s$ considered as $3$) for MPC and MAMPC}
    \label{tab:sd_te_q_tank_all_full}
\end{table}

\begin{table}[]
    \centering
    \begin{tabular}{|c|c|c|c|}
    \hline
       \multicolumn{1}{|p{2cm}|}{Method} & \multicolumn{1}{|p{2cm}|}{Sparse density $(\vartheta_{u_1})$} & \multicolumn{1}{|p{2cm}|}{Sparse density $(\vartheta_{u_2})$} & \multicolumn{1}{|p{2cm}|} {Tracking error $(\epsilon)$} \\
        \hline
        MAMPC ($n_s=1$) & 0.153 & 0.153 & 0.040  \\
        MAMPC ($n_s=3$) & 0.115 & 0.115 & 0.054 \\
        MPC & 0.553 & 0.584 & 0.237\\
        \hline
    \end{tabular}
    \caption{Quadruple tank system: Comparison of sparse density for a threshold value of $0.1$ and tracking errors (for a fixed value of $s$ considered as $3$)for MPC and MAMPC when the initial transient dynamics are not considered.}
    \label{tab:sd_te_q_tank_all}
\end{table}

It is evident from Table~\ref{tab:sd_te_q_tank_all_full} the input component $u1^3_{MAMPC}$  varied $19.3$ percent compared to 
$u1_{MPC}$ and $u1^1_{MAMPC}$  which varied  $57.9$ and $ 24.1$ percent respectively, which is a significant improvement when considered in an industrial setting, as it 
reduces the wear and tear of the control valves and the costs incurred with auxiliary equipment. In the case of input component $u2^3_{MAMPC}$, the control has been idle (within the threshold) for a $42$ and $4.8$ percent additionally over the optimal 
$u1_{MPC}$ and $u1^1_{MAMPC}$ respectively.
It is observed that the full tracking error is higher for MAMPC since the optimization is carried over a smaller 
feasible space and the initial error is also significant. 
To obtain a better picture of the non-transient errors, the initial transient dynamics are ignored by 
dropping a few time steps to calculate the tracking error and sparse density values. 
Table~\ref{tab:sd_te_q_tank_all} summarizes the results by dropping the first $15$ time steps from the 
results. However, the sparse density is still satisfactory and the tracking errors have been drastically 
reduced for MAMPC with $n_s=3$.

\subsection{Case study 2 : Solid Oxide Fuel Cell (SOFC)}
We consider the SOFC system, which has very fast dynamics compared to the Quadruple Tank system in the previous section. 
The identification of SOFC is based on the non-isothermal unchoked lumped
model given in~\cite{4410564}, which is briefly described in this
section. The considered model has three inputs (molar flow rate of fuel
($H_2+H_20$), air ($N_2+O_2$) and current ($I$)), four states (pressures of
$H_2,H_20,N_2 ~ \text{and} ~ O_2$)  and one output (voltage, $V$). The output equation is given by
\begin{equation*}
V=E-\eta_{\text{ohm}}-\eta_{\text{conc}}-\eta_{\text{act}}
\label{eq:fuelcellobseq}
\tag{12}
\end{equation*}
where
\begin{equation*}
\begin{split}
	\text{open circuit potential},~ E&=\frac{N_o}{2F}\Bigg{[}-{\Delta}g_f^o +RT~ln{\Bigg{(}}\frac{P_{H_2}  P_{O_2}^{0.5}}{P_{H_2O}  P_{\text{atm}}^{0.5}}{\Bigg{)}} \Bigg{]}\\
	\text{ohmic loss},~\eta_{\text{ohm}}&=r.I,\\
	\text{concentration loss},~\eta_{\text{conc}}&=-\frac{RT}{4F}\ln{\Bigg{(}}1-\frac{j}{j_L} \Bigg{)}\\
	\text{activation loss},~\eta_{\text{act}}&=\begin{cases}    \frac{RT}{4F}\frac{j}{j_0},& j\leq j_0\\
		\frac{RT}{2F}\text{ln}{\Big{(}}\frac{j}{j_0} \Big{)} +\frac{RT}{4F}\frac{j}{j_0}, & j > j_0
	\end{cases}
\end{split}
\end{equation*}
Here, ${\Delta}g_f^o$ is the change in molar specific Gibbs free energy of formation for the fuel cell reaction and is given
as $-{\Delta}g_f^o=188600 -56 (T - 1073.15)$. $r$ is the resistance, defined as $r=0.2\times{\text{exp}}{\Bigg{[}}-2870{\Bigg{(}}\frac{1}{1196.15}-\frac{1}{T} {\Bigg{)}} {\Bigg{]}}$. $T$ is the operating temperature of
SOFC, $j$ is the current density, which is equal to stack current divided by
area $(j=I/A)$, $j_L$ is the limiting current density and $j_0$ is the
exchange current density.

Considering the species balance of $H_2$, $H_2O$, $O_2$ and $N_2$, the state equations are obtained as
\begin{equation*}
\begin{split}
    \frac{dP_{H_2}}{dt}&= \frac{RT}{V_{an}}(\dot{n}_{H_2}^{\text{in}}-\dot{n}_{H_2}^{\text{out}}-2K_rI)\\
	\frac{dP_{H_2O}}{dt}&=  \frac{RT}{V_{an}}(\dot{n}_{H_2O}^{\text{in}}-\dot{n}_{H_2O}^{\text{out}}+2K_rI)\\
	\frac{dP_{O_2}}{dt}&= \frac{RT}{V_{cat}}(\dot{n}_{O_2}^{\text{in}}-\dot{n}_{O_2}^{\text{out}}-K_rI)\\
	\frac{dP_{N_2}}{dt}&= \frac{RT}{V_{cat}}(\dot{n}_{N_2}^{\text{in}}-\dot{n}_{N_2}^{\text{out}})
\end{split}
\label{eq:sofcsysdyn}
\tag{13}
\end{equation*}
where, $K_r=N_o/4F$ , $I$ is the stack current, $N_o$ is number of cells in the stack, $F$ is the Faraday's constant,
$\dot{n}_{H_2}^{\text{in}}$ is the inlet molar flow rate of $H_2$,
$\dot{n}_{H_2}^{\text{out}}$ is outlet molar flow rate of $H_2$ and $P_{H_2}$ is
the partial pressure of $H_2$ in the stack. Similar definitions hold for the
remaining components.

The outlet molar flow rates are given by the following equations 
\begin{equation*}
\begin{split}
    \dot{n}_{H_2}^{\text{out}}&=C{A_a}{P_{H_2}}\sqrt{\frac{2(P_{H_2}+P_{H_2O}-P_{\text{atm}})}{RT(P_{H_2}M_{H_2}+P_{H_2O}M_{H_2O})}}\\
	\dot{n}_{H_2O}^{\text{out}}&=C{A_a}{P_{H_2O}}\sqrt{\frac{2(P_{H_2}+P_{H_2O}-P_{\text{atm}})}{RT(P_{H_2}M_{H_2}+P_{H_2O}M_{H_2O})}}\\
	\dot{n}_{O_2}^{\text{out}}&=C{A_c}{P_{O_2}}\sqrt{\frac{2(P_{O_2}+P_{N_2}-P_{\text{atm}})}{RT(P_{O_2}M_{O_2}+P_{N_2}M_{N_2})}}\\
	\dot{n}_{N_2}^{\text{out}}&=C{A_c}{P_{N_2}}\sqrt{\frac{2(P_{O_2}+P_{N_2}-P_{\text{atm}})}{RT(P_{O_2}M_{O_2}+P_{N_2}M_{N_2})}}\\
\end{split}	
\end{equation*}
where $M_c$ is the molecular weight of the component $c$. $A_a$, $A_c$ are
anode side and cathode side cross-sectional areas
respectively. $C=\frac{C_d}{\sqrt{(1-(D_2/D_1)^4)}}$ where $C_d$ is the
discharge coefficient of orifice, $D_2$ and $D_1$ are diameters of the orifice and
manifold respectively.

\begin{table}[]
	\centering
        \caption{Parameters of SOFC system}
        \begin{tabular}{|l|c|c|}
            \hline
            Parameter & Value  & Units  \\
            \hline
            No of cells, $N_o$ 					& 384 & - \\
            Anode cross sectional area, $A_a$ 	& 0.0025 & $m^2$\\
            Cathode cross sectional area, $A_c$ & 0.0025 & $m^2$ \\
            Coefficient of discharge, $C_d$ 	& 0.75 & - \\
            Temperature, $T$ 					& 1273.15 & $K$ \\
            Inlet fuel flow rate 				& 1.2 & $mol/s$\\
            Inlet air flow rate 				& 5 & $mol/s$ \\
            Limiting current density, $j_L$ 	& 1500 & $A/m^2$ \\
            Exchange current density, $j_o$ 	& 10000 & $A/m^2$ \\
            Current, $I$                        & 400 & $A$\\
            Faraday's constant, $F$             & 96485 & $C/mol$\\
            \hline
        \end{tabular}%
        \label{parm_sofc}
\end{table}
\subsubsection{Simulation}
\begin{figure}[]
    \centering    \includegraphics[width=0.5\linewidth]{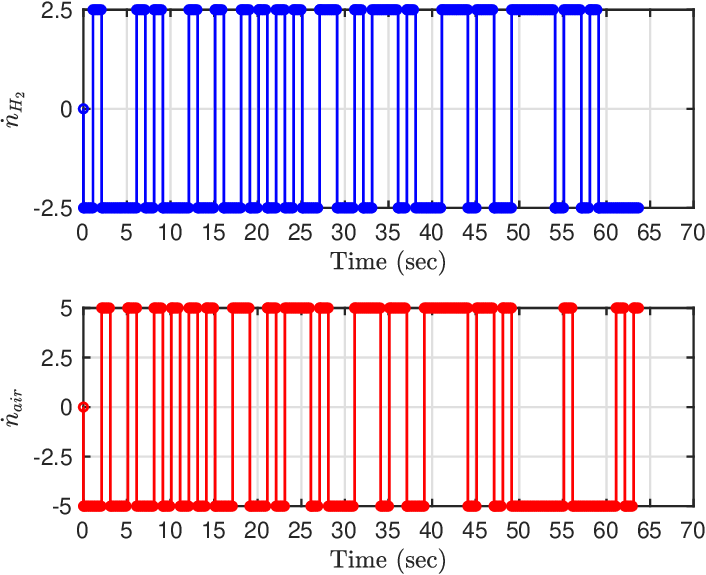}
    \caption{PRBS as input for $\dot{n}_{H_2}$ and $\dot{n}_{air}$ in fuel cell stack system.}
    \label{fig:sofc_prbs_u}
\end{figure}

\begin{figure}[]
    \centering
    \includegraphics[width=0.5\linewidth]{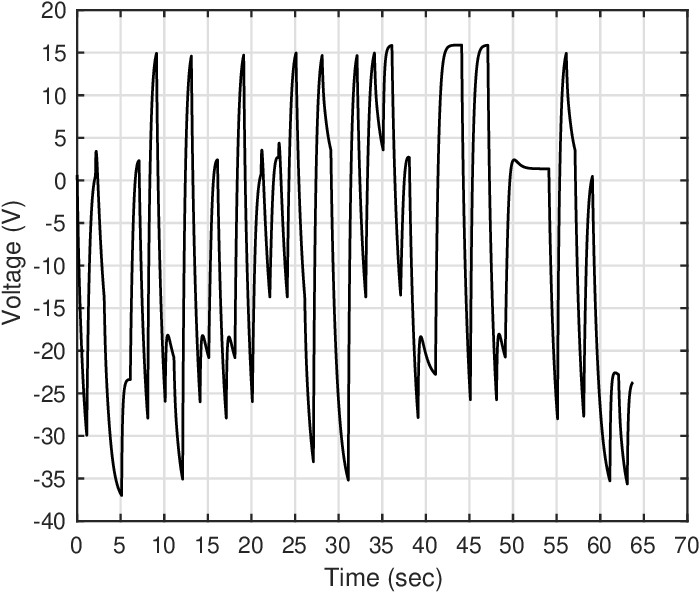}
    \caption{Output voltage of fuel cell stack for the above considered PRBS inputs}
    \label{fig:sofc_sim_y}
\end{figure}

The dynamic model explained above is simulated in the MATLAB-Simulink\textsuperscript{\textregistered}   environment with the
parameters given in Table~\ref{parm_sofc}. In simulation, a PRBS signal is used as input to excite multiple 
frequencies in a system. Initially, the system is brought to a steady state, and then the PRBS signal is fed to 
each of the inputs of the fuel cell stack i.e., Fuel flow rate $(\dot{n}_{H_2})$ and Air flow rate 
$(\dot{n}_{air})$. A perturbation magnitude of $5$ and $10$ were used for each of these inputs, which can be 
seen in Figure~\ref{fig:sofc_prbs_u}. The output voltage can be observed to oscillate in a narrow band of 
$-40$ to $20$ referred from Figure~\ref{fig:sofc_sim_y}. It is important to note in the simulation plots, the 
steady state part is removed, which is a pre-processing step in the identification exercise. 

\subsubsection{Identification}
A deterministic model is identified from the simulated input-output data using N4SID~\cite{van1994n4sid} subspace identification algorithm. The identified model is used in optimal control design, and the results are given below. 
\subsubsection{Comparison of MPC and MAMPC Results}
\begin{figure}
    \centering
    \includegraphics[width=0.5\linewidth]{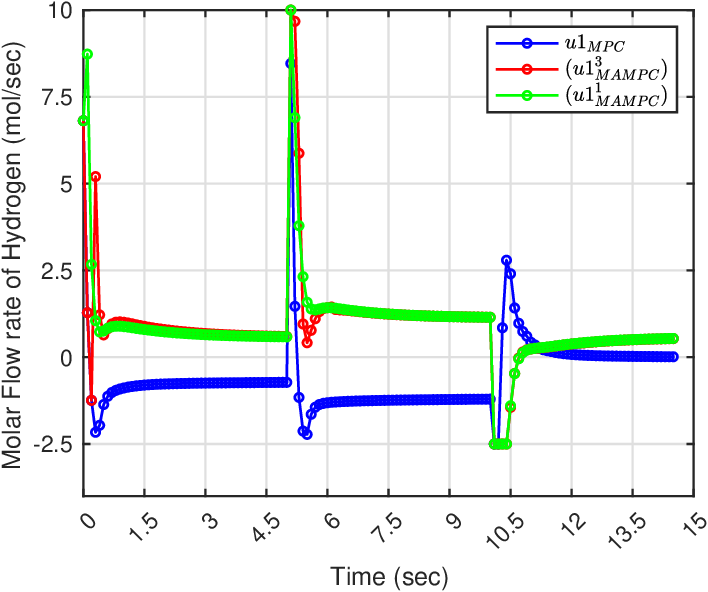}
    \caption{Optimal profile of  manipulated variable i.e, molar flowrate of $\dot{n}_{H_2}$ for the fuel cell stack system given by MPC and MAMPC (two cases are considered with $n_s=1$ and $n_s=3$ as before and $s = 3$)}
    \label{fig:sofc_u1_all}
\end{figure}

\begin{figure}
    \centering
    \includegraphics[width=0.5\linewidth]{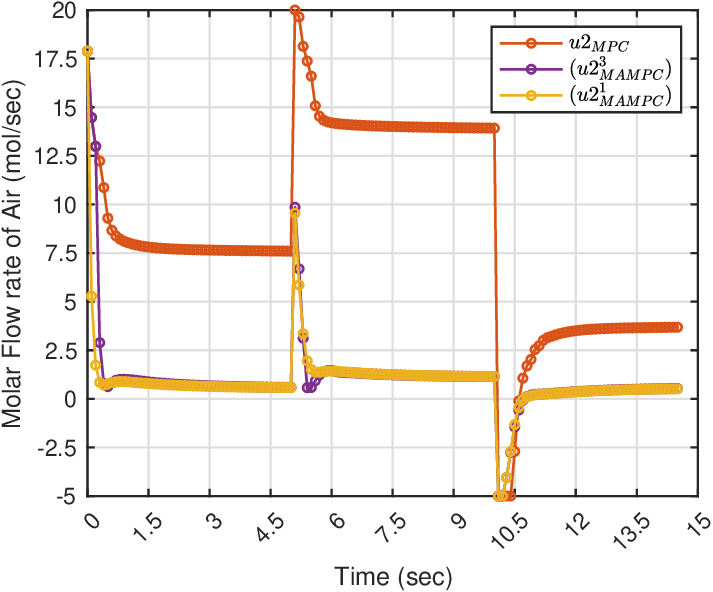}
    \caption{Optimal profile of  manipulated variable i.e, molar flowrate of $\dot{n}_{Air}$ for the fuel cell stack system given by MPC and MAMPC (two cases are considered with $n_s=1$ and $n_s=3$ as before and $s = 3$)}
    \label{fig:sofc_u2_all}
\end{figure}

\begin{figure}
    \centering
    \includegraphics[width=0.5\linewidth]{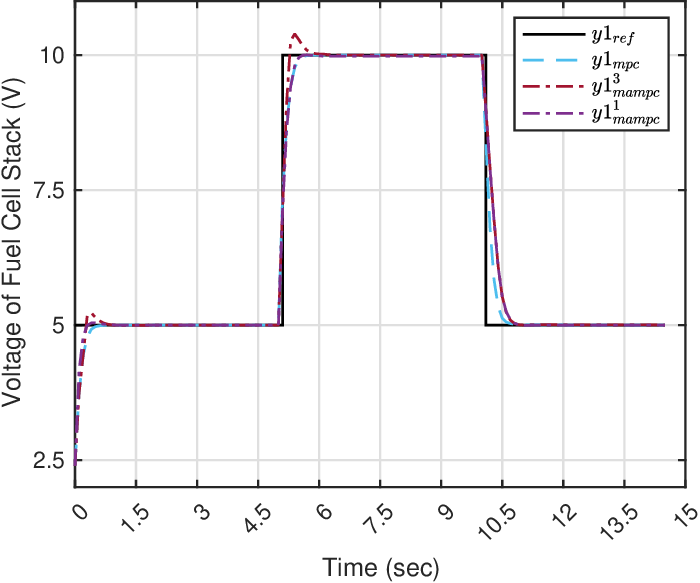}
    \caption{Output voltage of fuel cell stack for the given reference for the optimal input profiles given by MPC and MAMPC (two cases are considered with $n_s=1$ and $n_s=3$ as before and $s = 3$)}
    \label{fig:sofc_y1_all}
\end{figure}

The same design conditions are used as mentioned for the quadruple tank systems for the control of fuel cell stack voltage using MPC and MAMPC formulations. The optimal input profiles and the closed loop responses of both the methods can be referred from Figures~\ref{fig:sofc_u1_all},~\ref{fig:sofc_u2_all} and ~\ref{fig:sofc_y1_all}.

Although the optimal input profiles of $u1_{MPC}$ are smooth, it is evident from Table~\ref{tab:sd_te_sofc_all_full} that $u1^1_{MAMPC}$ varied $10.3\%$ compared to $u1^3_{MAMPC}$ (13.7\%)  and $u1_{MPC}$ (15.1\%) respectively. It is evident from the results that $u1^1_{MAMPC}$ and $u1^3_{MAMPC}$ remain within the threshold for an additional period $4.8\%$ and $1.4\%$ more than $u1_{MPC}$ respectively. Similarly for the input $u2$, $u2^1_{MAMPC}$ and $u2^3_{MAMPC}$ are within the threshold and $6.9\%$, $2.7\%$ are the improvement factors over $u2_{MPC}$.
As evident from Tables~\ref{tab:sd_te_sofc_all_full} and \ref{tab:sd_te_sofc_all}, the tracking errors for 
$u1^3_{MAMPC}$ and $u1^1_{MAMPC}$ are higher compared to $u1_{MPC}$.

\begin{table}[]
    \centering
    \begin{tabular}{|c|c|c|c|c|}
    \hline
       \multicolumn{1}{|p{2cm}|}{Method} & \multicolumn{1}{|p{2cm}|}{Sparse density $(\vartheta_{u_1})$} & \multicolumn{1}{|p{2cm}|}{Sparse density $(\vartheta_{u_2})$} & \multicolumn{1}{|p{2cm}|} {Tracking error $(\epsilon)$}\\
        \hline
        MAMPC ($n_s=1$) & 0.103 & 0.117  & 0.326 \\
        MAMPC ($n_s=3$) & 0.137 & 0.144  & 0.327\\
        MPC & 0.151 & 0.186 &  0.227\\
        \hline
    \end{tabular}
    \caption{SOFC System : Comparison of sparse density for a threshold value of $0.1$ and their respective tracking errors (for a fixed value of $s$ considered as $3$) for MPC and MAMPC}
    \label{tab:sd_te_sofc_all_full}
\end{table}

\begin{table}[]
    \centering
    \begin{tabular}{|c|c|c|c|c|}
    \hline
       \multicolumn{1}{|p{2cm}|}{Method} & \multicolumn{1}{|p{2cm}|}{Sparse density $(\vartheta_{u_1})$} & \multicolumn{1}{|p{2cm}|}{Sparse density $(\vartheta_{u_2})$} & \multicolumn{1}{|p{2cm}|} {Tracking error $(\epsilon)$} \\
        \hline
        MAMPC ($n_s=1$) & 0.084 & 0.098  & 0.280 \\
        MAMPC ($n_s=3$) & 0.119 & 0.126  & 0.273 \\
        MPC & 0.133 & 0.169 & 0.171\\
        \hline
    \end{tabular}
    \caption{SOFC System: Comparison of sparse density for a threshold value of $0.1$ and their respective tracking errors (for a fixed value of $s$ considered as $3$) for MPC and MAMPC when the initial transient dynamics is not considered.}
    \label{tab:sd_te_sofc_all}
\end{table}

\section{Conclusion}
\label{sec:conclusion}
In this paper, a control framework for the receding horizon approach is considered to possess 
the minimum attention property, which is called MAMPC. An alternating minimization algorithm is proposed to 
solve the MAMPC problem and is extensively studied using examples from the control 
literature.
The novelty of the paper lies in the MAMPC problem framework, which considers the previous optimal inputs in 
the optimization framework using a sparsity horizon. An alternating minimization approach to solve the 
optimization problem with zero norm constraint has been proposed. The minimum attention property is ensured 
in the receding horizon framework of MPC, and the resulting control inputs possess the minimum 
attention property, i.e., the input infrequently changes over the horizon. A detailed comparison with respect 
to MPC has been provided. The effectiveness of our method is studied using the sparse density, and the 
calculated values are reported.

\bibliographystyle{plain}
\bibliography{myref}

\end{document}